\documentclass[apjl]{emulateapj}

\usepackage{apjfonts}
\usepackage{graphicx}
\usepackage{natbib}
\usepackage{subfigure}

\newcommand{\lsi}{LS~I~+61\degr303}
\newcommand{\fermi}{{\em Fermi}}
\newcommand{\hess}{H.E.S.S.}

\shorttitle{{\em FERMI} OBSERVATIONS OF LS 5039}
\shortauthors{ABDO ET AL.}

\begin{document}

\title{{\em Fermi} LAT Observations of LS 5039}

\author{
A.~A.~Abdo\altaffilmark{1,2}, 
M.~Ackermann\altaffilmark{3}, 
M.~Ajello\altaffilmark{3}, 
W.~B.~Atwood\altaffilmark{4}, 
M.~Axelsson\altaffilmark{5,6}, 
L.~Baldini\altaffilmark{7}, 
J.~Ballet\altaffilmark{8}, 
G.~Barbiellini\altaffilmark{9,10}, 
D.~Bastieri\altaffilmark{11,12}, 
B.~M.~Baughman\altaffilmark{13}
% ...
}

\altaffiltext{1}{Space Science Division, Naval Research Laboratory, Washington, DC 20375, USA}
\altaffiltext{2}{National Research Council Research Associate, National Academy of Sciences, Washington, DC 20001, USA}
\altaffiltext{3}{W. W. Hansen Experimental Physics Laboratory, Kavli Institute for Particle Astrophysics and Cosmology, Department of Physics and SLAC National Accelerator Laboratory, Stanford University, Stanford, CA 94305, USA}
\altaffiltext{4}{Santa Cruz Institute for Particle Physics, Department of Physics and Department of Astronomy and Astrophysics, University of California at Santa Cruz, Santa Cruz, CA 95064, USA}
\altaffiltext{5}{Department of Astronomy, Stockholm University, SE-106 91 Stockholm, Sweden}
\altaffiltext{6}{The Oskar Klein Centre for Cosmoparticle Physics, AlbaNova, SE-106 91 Stockholm, Sweden}
\altaffiltext{7}{Istituto Nazionale di Fisica Nucleare, Sezione di Pisa, I-56127 Pisa, Italy}
\altaffiltext{8}{Laboratoire AIM, CEA-IRFU/CNRS/Universit\'e Paris Diderot, Service d'Astrophysique, CEA Saclay, 91191 Gif sur Yvette, France}
\altaffiltext{9}{Istituto Nazionale di Fisica Nucleare, Sezione di Trieste, I-34127 Trieste, Italy}
\altaffiltext{10}{Dipartimento di Fisica, Universit\`a di Trieste, I-34127 Trieste, Italy}
\altaffiltext{11}{Istituto Nazionale di Fisica Nucleare, Sezione di Padova, I-35131 Padova, Italy}
\altaffiltext{12}{Dipartimento di Fisica ``G. Galilei", Universit\`a di Padova, I-35131 Padova, Italy}
\altaffiltext{13}{Department of Physics, Center for Cosmology and Astro-Particle Physics, The Ohio State University, Columbus, OH 43210, USA}
% ...

\begin{abstract}
The first results from observations of the high mass X-ray binary LS
5039 using the \fermi\ Gamma-ray Space Telescope data between 2008
August and 2009 June are presented.  Our results indicate variability that is
consistent with the binary period, with the emission being modulated
with a period of $3.903 \pm 0.005$ days; the first detection of this
modulation at GeV energies.  The light curve is characterized by a
broad peak around superior conjunction in agreement with inverse
Compton scattering models.  The spectrum is represented by a power law
with an exponential cutoff, yielding an overall flux (100 MeV -- 300
GeV) of 4.9 $\pm$ 0.5(stat) $\pm$ 1.8(syst) $\times$10$^{-7}$ photon
cm$^{-2}$ s$^{-1}$, with a cutoff at 2.1 $\pm$ 0.3(stat) $\pm$
1.1(syst) GeV and photon index $\Gamma$ = 1.9 $\pm$ 0.1(stat) $\pm$
0.3(syst).  The spectrum is observed to vary with orbital phase,
specifically between inferior and superior conjunction.  We suggest
that the presence of a cutoff in the spectrum may be indicative of
magnetospheric emission similar to the emission seen in many pulsars
by {\em Fermi}.
\end{abstract}

\section{Introduction}

LS 5039, PSR B1259$-$63 and \lsi\ are the only binaries, with
high-mass companions, long known to be spatially coincident with
sources detected at energies greater than 100 MeV, e.g., those listed
in the Third Energetic Gamma-Ray Experiment (EGRET) catalog
\citep{1999ApJS..123...79H}. The latter binary was detected by the
Large Area Telescope (LAT) on the \fermi\ mission, confirming it as a
GeV gamma-ray source and finding variability that is consistent with
modulation on the binary period of 26.6 $\pm$ 0.5 days
\citep{2009ApJ...701L.123A}. This constituted the first detection of
orbital periodicity in high-energy (HE) gamma rays (20 MeV -- 100
GeV). In this Letter, we present the results of \fermi\ observations
of LS 5039.

LS 5039 is one of a handful of X-ray binaries that have been detected
recently at very HE $\gamma$-rays; results from $\sim70$ hr of
observations distributed over many orbital cycles have been presented
by \citet{Aharonian:2005nj, Aharonian:2006qw}.  These observations
yielded a modulation of the very high energy (VHE, >100 GeV) gamma-ray
flux with a period of 3.9078$\pm$0.0015 days \citep{Aharonian:2006qw},
consistent with the orbital period as determined by
\citet{Casares:2005gg} from optical spectroscopy. Short timescale
variability displayed on top of this periodic behavior, both in flux
and spectrum, was also reported.

The nature of the LS 5039 compact object is unknown: a black hole and
a neutron star have been invoked as possible compact object companions
in a slightly eccentric ($e\sim0.35$), 3.90603$\pm$0.00017 day orbit
around the O6.5V star \citep{Casares:2005gg}. The discovery of a
jet-like radio structure in LS 5039 coincident with an X-ray and EGRET
source prompted a microquasar interpretation
\citep{2000Sci...288.2340P}. Variability in the EGRET source could not
be established precisely \citep{2001A&A...370..468T,
  2003ApJ...597..615N}. Recently it has been shown that the X-ray flux
is modulated on the orbital period \citep{Takahashi:2008vu}.

\citet{2008A&A...481...17R} provided Very Long Baseline Array (VLBA)
radio observations of LS 5039 with morphological and astrometric
information at milliarcsecond scales that cannot easily be explained
by a microquasar scenario. \citet{2005A&A...430..245M} assessed the
low X-ray state, showing the absence of accretion features in the
X-ray spectra.  Thus, measurements at radio and VHE $\gamma$-rays in
the cases of \lsi\ \citep{2006smqw.confE..52D, 2008ApJ...684.1351A} or
PSR B1259$-$63 \citep{2005A&A...442....1A}, whose overall spectral
energy distribution is similar to that of LS 5039, gave the impression
that all three systems are different realizations of the same
scenario: a pulsar-massive star binary \citep[][]{Dubus:2006lc}.

Theoretical computations of the gamma-ray emission in both compact object 
scenarios have been made, with gamma-rays produced by
inverse Compton (IC) scattering of the stellar light by VHE electrons
accelerated in the vicinity of the compact object. In the case of the
black hole companion, HE and VHE emission would result from particles
accelerated in the jet \citep[][]{2007A&A...464..259B,
  2007APh....27..278B, 2008MNRAS.383..467K}.
Alternatively, it would involve the relativistic wind of a young,
rotation-powered pulsar, either as a result of particle acceleration
in the wind interaction region \citep{Dubus:2006lc} or by processes
within the pulsar wind \citep{2008ApJ...674L..89S, 2008APh....30..239S,
2008A&A...488...37C}.

\begin{figure*}
  \centering
   \includegraphics[width=0.9\linewidth,angle=0,clip]{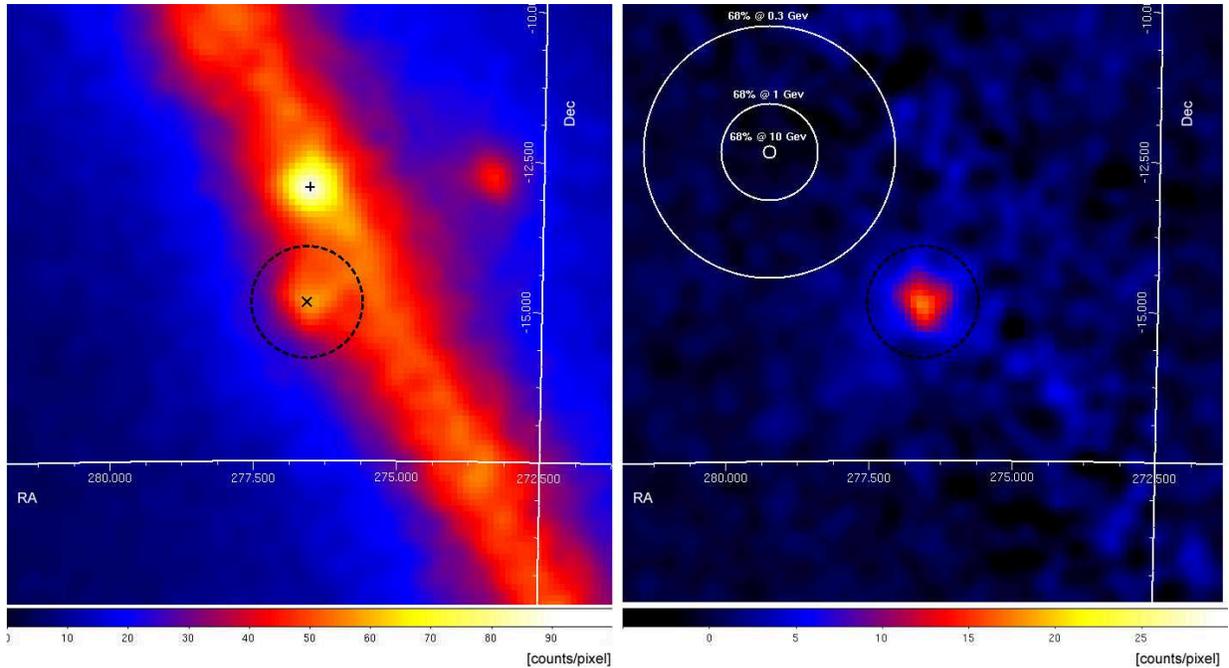}
\caption{Left: the smoothed counts map for 100~MeV--300~GeV of a
  10\degr $\times$10\degr\ region around the LS 5039 location (marked
  with ``$\times$''); the dashed black circle indicates the 0.$^{\circ}$925
  timing analysis aperture.  A 0.$^\circ$3 gaussian smoothing function
  was applied to the 0.$^\circ$1 bins. PSR J1826$-$1256 is marked with
   ``+''. The exposure varies by less than 7\% across the field at a
  representative energy of 10 GeV. Right: residuals of left
  panel after subtracting all modelled sources excluding LS 5039 and
  excluding events which arrive during the peaks in the pulsar phase
  cycle of PSR J1826$-$1256 (see section~\ref{sec:spectra}). The white
  circles indicate the 68\% LAT containment region at three energies: 0.3,
  1, and 10 GeV.  Note that the color scales are different for the two
  panels.}
\label{CountsMap}
\end{figure*}

\section{Data Reduction and results}

The LAT onboard \fermi\ is an electron--positron
pair production telescope, featuring solid state silicon trackers and
cesium iodide calorimeters, sensitive to photons from $\sim 20$ MeV to
$>300$ GeV \citep{2009ApJ...697.1071A}.

The analysis dataset spanned 2008 August 4, through 2009
June 22. The data were reduced and analyzed using the {\em Fermi} Science Tools
v9r15 package\footnote{See the FSSC
  website for details of the Science Tools:
  http://fermi.gsfc.nasa.gov/ssc/data/analysis/}. The standard onboard
filtering, event reconstruction, and classification were applied to
the data \citep{2009ApJ...697.1071A}, and for this analysis the
high-quality (``Pass 6 diffuse") event class is used.  Throughout the
analysis, the ``Pass 6 v3 Diffuse'' (P6\_V3\_DIFFUSE) instrument
response functions (IRFs) are applied.

Time periods when the region around LS 5039 was observed at a zenith
angle greater than 105\degr\ and for observatory rocking angles of
greater than 43$^\circ$ were also excluded to avoid contamination from
the Earth albedo photons. With these cuts, a photon count map of a
10$^\circ$ region around the binary is shown in Figure~\ref{CountsMap}.

LS 5039 is detected at a level of 28.5$\sigma$ (see section~\ref{sec:spectra}).
The {\tt gtfindsrc} tool finds a best-fit location for LS 5039 of
R.A.~=~18$^{\rm h}$26$^{\rm m}$24\fs7,
decl.~=~$-$14\degr48\arcmin39\arcsec (J2000) with a 95\% error of
0.054\degr (including a 20\% systematic error derived from the internal \fermi\ 11-month catalog).  The nominal position of LS 5039 at R.A.~=~18$^{\rm
  h}$26$^{\rm m}$15\fs1, Dec.~=~$-$14\degr50\arcmin54.\arcsec3 (J2000)
\citep{2004AJ....127.3043Z} lies just on the \fermi\ 95\%
contour; the nominal position was used throughout the analysis.

\section{Timing Analysis}

LAT light curves were extracted using aperture photometry. The LAT
point-spread function (PSF) is strongly energy dependent and, particularly
since LS 5039 is located in the Galactic plane, there is also
significant contribution to flux within an aperture from diffuse
emission and point sources that depends on the aperture size and the
energy range used.  The aperture and energy band employed were
independently chosen to maximize the signal-to-noise level. The
optimum aperture radius was found to be approximately 0.\degr925 in
the energy range 100 MeV--10 GeV. The time binning of the light curve
was 1000 s.  Exposures were calculated using {\tt gtexposure} assuming
a power-law spectrum with a photon index of $\Gamma$~=~2.5.

\begin{figure}[t]
\includegraphics[width=0.9\linewidth,clip]{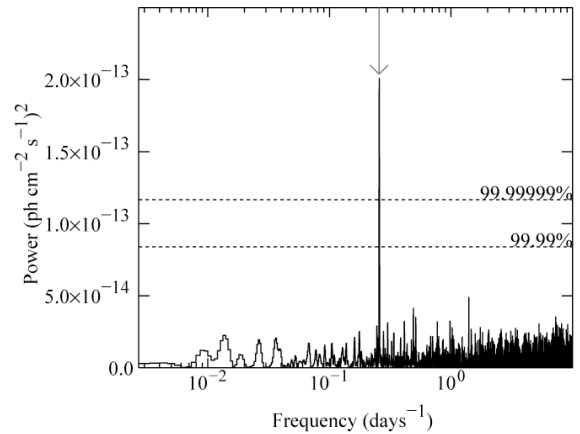}
\caption{Power spectrum of the light curve. The arrow indicates the
  known orbital period of 3.90603 days \citep{Casares:2005gg}.  The
  dashed lines show the 99.99\% and 99.99999\% significance levels.}
\label{Powspec}
\end{figure}

A search was made for periodic modulation by calculating the weighted
periodogram of the light curve \citep{1976Ap&SS..39..447L,
  1982ApJ...263..835S, 2007AIPC..921..548C}.  Since the exposure of
the time bins was variable, the contribution of each time bin to the
power spectrum was weighted based on its relative exposure. The
periodogram is shown in Figure~\ref{Powspec}. The arrow marks the
\citet{Casares:2005gg} orbital period and a highly significant peak is
detected at this period; the false alarm probability is
$\sim$10$^{-10}$.  The significance levels marked are for a ``blind''
search with 5000 independent frequency steps, however, the effects of
the tuning of the aperture radius and energy range are not taken into
account.  The period error was estimated using a Monte Carlo approach:
light curves were simulated using the observed LS 5039 light curve and
randomly shuffling the data points within their errors, assuming
Gaussian statistics. The corresponding periodogram was then calculated
and the location of the peak at $\sim$3.9 days recorded. From
$\sim$200,000 simulations, we calculate an error estimate of the
orbital period of 3.903 $\pm$ 0.005 days (1$\sigma$).

\section{Spectral Analysis}\label{sec:spectra}

The {\tt gtlike} likelihood fitting tool was used to perform the
spectral analysis, wherein a spectral-spatial model containing
  point and diffuse sources is created and the parameters obtained
  from a simultaneous maximum likelihood fit to the data.  The
10\degr\ region around the source was modeled for Galactic and
isotropic diffuse contributions and 22 additional significant
point sources (taken from the internal \fermi\ 11-month catalog) were
included; point sources were modeled with simple power laws with the
exception of the bright, nearby pulsar (PSR J1826$-$1256) for which a
power law with an exponential cut-off was used.  The flux contribution
of PSR J1826$-$1256 to the region was minimized by excluding events
which arrive during the peaks in the pulsar phase cycle \citep[
  0.175$< \phi <$0.3 and 0.625$< \phi <$0.775 were excluded;
  see][]{2009Sci...325..840A}. A scaling factor of 1/0.725 is applied
to measured fluxes to account for livetime loss due to this cut.

The 10\degr\ region was chosen to capture the broad PSF obtained at
100 MeV. An alternate fitting method using energy-dependent regions of
interest was used, yielding compatible results that were folded into
the systematic errors.

In analyzing the emission of LS 5039, we used models for the Galactic
diffuse emission ($gll\_iem\_v{\it 02}.fit$) and isotropic backgrounds
currently recommended by the LAT team\footnote{Descriptions of the
  models are available from the FSSC: http://fermi.gsfc.nasa.gov/}.
The model for the Galactic diffuse emission was developed using
spectral line surveys of $\mbox{H\,{\sc i}}$, CO (as a tracer of
$\mbox{H}_{2}$) to derive the distribution of interstellar gas in
Galactocentric rings.

The model of the diffuse gamma-ray emission was then
constructed by fitting the gamma-ray emissivities of the rings in
several energy bands to the LAT observations.  The fitting also
required a model of the IC emission calculated using
GALPROP \citep{2007Ap&SS.309...35S} and a model for the isotropic
diffuse emission.  The latter was fitted to the LAT data using an
analysis of the sky above $|b|$ = 30 $ ^\circ$ and includes the
significant contribution of residual (misclassified) cosmic rays at
HEs for the current IRFs.
  
\begin{figure}[t]
 \includegraphics[width=0.9\linewidth,angle=0,clip]{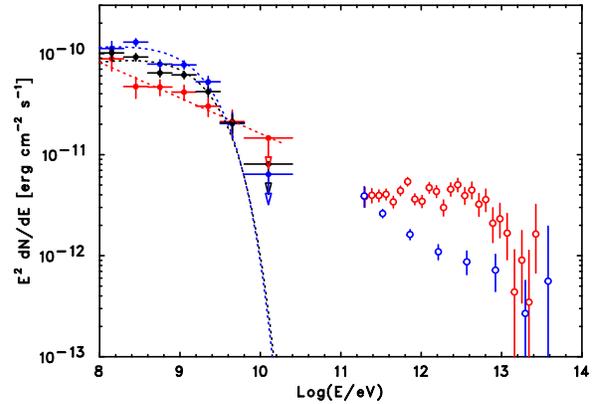}
\caption{Fitted spectrum of LS 5039. Fermi data points are from
  likelihood fits in each energy bin. The black points (dotted line)
  represent the phase-averaged {\em Fermi}/LAT spectrum.  The red data
  points (dotted line) represent the spectrum (overall fit) at
  inferior conjunction (Phase 0.45--0.9); blue data points (dotted
  line) represent the spectrum (overall fit) at superior conjunction
  (Phases, $<$0.45 and $>$0.9).  Data points above 100 GeV are taken
  from \hess\ observations \citep{Aharonian:2006qw}; the
  data from \hess\ are not contemporaneous with \fermi, though they do
  cover multiple orbital periods.}
\label{PhaseAveFit}
\end{figure}

\begin{figure}[t]
 \includegraphics[width=0.9\linewidth,angle=0,clip]{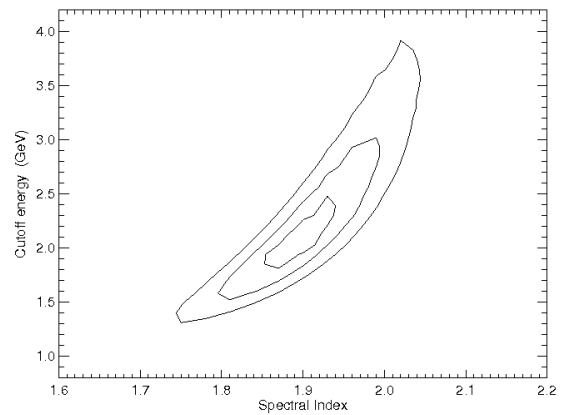}
\caption{The 1$\sigma$, 2$\sigma$, and 3$\sigma$ contours for the photon index and cutoff energy spectral parameters from fitting to the phase-averaged \fermi\ data.}
\label{contours}
\end{figure}

Initially a simple power law, $E^{-\Gamma}$, was fitted to the data from
all orbital phases yielding a photon index of $\Gamma \sim 2.54$.
However, as clearly seen in Figure~\ref{PhaseAveFit}, the energy
spectrum appears to turn over at energies above $\sim 2$~GeV.  The
possibility of an exponential cutoff was investigated in the form
$E^{-\Gamma} \exp{[-(E/E_{\rm cutoff})]}$. The chance probability to
incorrectly reject the power-law hypothesis was found to be 1.6
$\times$10$^{-16}$.  The maximum likelihood exponential cutoff
spectral model has a likelihood test statistic
\citep{1996ApJ...461..396M} value of $\sim$814.6, or
28.5$\sigma$.  The photon index is $\Gamma =1.9 \pm 0.1\ ({\rm stat})
\pm 0.3 ({\rm syst})$; the 100 MeV-- 300 GeV flux is $(4.9 \pm
0.5\ ({\rm stat}) \pm 1.8 \ ({\rm syst})) \times 10^{-7} ~{\rm
  ph}~{\rm cm}^{-2}~{\rm s}^{-1}$ and the cutoff energy is $2.1 \pm
0.3\ ({\rm stat}) \pm 1.1 ({\rm syst})~{\rm GeV}$ (see below for a
discussion of systematics).  The correlation beween the photon index
and cutoff energy was explored by fitting a family of models over a
grid of indices and cutoff energies centered on the best-fit parameter
values.  The 1, 2 and 3$\sigma$ error contours are shown in
Figure~\ref{contours}; both parameters are bounded and well
constrained. A total of 359,789 photons were found in the
10\degr\ region. Evaluating the fit parameters, 3992 $\pm$ 63 photons
were observed from LS 5039 above 100 MeV.

A number of effects are expected to contribute to the systematic
errors. The largest is uncertainty in the diffuse modeling as
evidenced by the intense swath of photons along the Galactic plane
shown in Fig ~\ref{CountsMap}. A reasonable range of shape difference
has been explored using the GALPROP model of the region. Both models
give reasonable residuals maps and show differences of 10\%, 37\%, and
50\%, respectively for index, cutoff energy and flux. In all diffuse
models tested, the exponential cutoff model is a significant
improvement over the power law.

 The impact of systematic uncertainties due to the IRFs are estimated
 by using outlier IRFs that bracket our nominal ones in effective
 area. These are defined by envelopes above and below the P6\_V3 IRFs
 by linearly connecting differences of (10\%, 5\%, and 20\%) at
 log($E/$MeV) of (2, 2.75, and 4) respectively.  The variation on the
 index was 15\%; the other parameter variations were small compared to
 those due to the diffuse modeling. Other effects considered are:
 fitting technique, cuts applied (minimum and maximum energies), but
 they are all within the ranges defined by diffuse modeling and
 bracketing effective area variations.

\begin{figure}[t]
\includegraphics[width=0.9\linewidth]{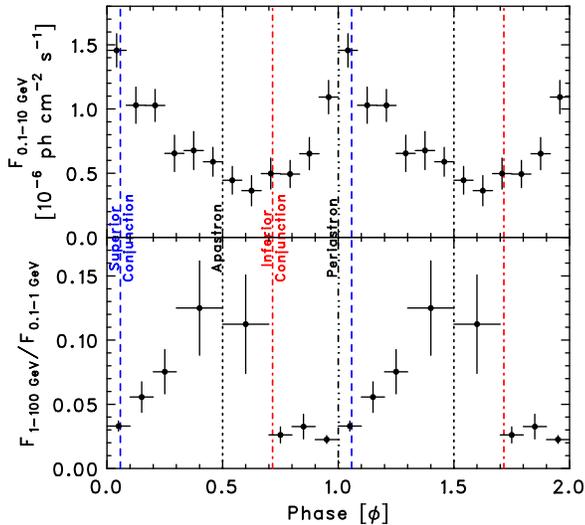}
\caption{Top: flux vs. orbital phase for 0.1-10 GeV. Bottom: variations with orbital phase in the hardness ratio of 1-100 to 0.1-1 GeV.}
\label{PhaseIndex}
\end{figure}
 
\subsection{Phase resolved spectral analysis}
We also searched for orbital dependence of the spectral shape.  {\tt
  gtlike} fits were performed for each phase interval of $\Delta \phi
= 0.1$ in the same way as for the phase-averaged spectral
analysis. Figure~\ref{PhaseIndex} shows the hardness ratio, $F_{1-100
  {\rm GeV}} /F_{0.1-1 {\rm GeV}}$, as a function of orbital phase.
Due to limited statistics, bin widths of $\Delta \phi$ of $0.2$
are used for the phase interval $\phi = 0.3$--0.6, while $\Delta \phi
= 0.1$ are employed for the other phase intervals. The spectral shape
varied such that the spectrum is softer around periastron and is
harder around apastron.

\citet{Aharonian:2006qw} define two broad phase intervals, inferior
conjunction (0.45$< \phi <$0.9) and superior conjunction (0.9 $< \phi
<$0.45), both \hess\ spectra being shown in Figure~\ref{PhaseAveFit}.
Taking the same phase intervals with the LAT data, we find a power-law
spectrum with $\Gamma = 2.25\pm0.11$ at inferior conjunction; an
energy cutoff was not statistically justified.
At superior conjunction a power law with an exponential cutoff was
preferred with $\Gamma = 1.91\pm0.16$ and a cutoff energy of
1.9$\pm$0.5 GeV.  \citet{Aharonian:2006qw} report spectral variability
in the source between superior and inferior conjunction, however, they
do not see any indication of long-term variability suggesting that it
may be reasonable to compare non-contemporaneous observations.

\section{Discussion}

The initial association of LS 5039 with the EGRET source 3EG
J1824$-$1514 \citep{Paredes:2000ci} had remained tentative due to the
large EGRET error circle and the lack of timing signatures. The
association was bolstered by the discovery of point-like, modulated
gamma-ray emission above 100 GeV
\citep{Aharonian:2005nj,Aharonian:2006qw}. The {\em Fermi}
observations enabled the detection of an orbital modulation,
indicating that the binary is also a source of gamma rays above 100
MeV. The periods determined independently from the {\em Fermi}-LAT and
\hess\ data are self consistent and compatible with the binary period
obtained from radial velocity measurements of the companion
\citep{2009ApJ...698..514A}. This is only the second high mass X-ray
binary after \lsi\ \citep{2009ApJ...701L.123A} to become a confirmed
emitter in the HE gamma-ray domain.

The short orbital period means that the compact object passes within a
stellar radius of the surface of the $kT_\star \approx 3$ eV, $R_\star
\approx 10 R_\odot$ companion \citep{Casares:2005gg}. Gamma rays
emitted in the vicinity of the compact object with energies above the
30 GeV threshold inevitably pair produce with stellar photons
\citep[see
  e.g.][]{1987ApJ...322..838P,1995SSRv...72..593M,Bottcher:2005zx,Dubus:2006lr}.
Emission in the {\em Fermi} range is largely unaffected by absorption
and should allow easier identification of the intrinsic spectrum and
variability of the source.  However, it can be affected by cascading
of higher-energy photons.

The HE modulation peaks at $\phi \sim$0.0--0.1, close to superior
conjunction ($\phi=0.06$) while the VHE modulation peaks close to
inferior conjunction ($\phi=0.72$). The phase difference can be
explained mostly as a result of the competition between IC scattering
on HE electrons and VHE pair production, assuming the
companion star to be close to the compact object. The star provides
target photons for both processes with the radiation density varying
by a factor 4 along the eccentric orbit. IC scattering will vary with
radiation density but, because the source of seed photons is
anisotropic, the flux will also depend on the geometry seen by the
observer in non-trivial ways
\citep{2008MNRAS.383..467K, 2008APh....30..239S}. The emission is
enhanced (reduced) when the highly relativistic electrons seen by the
observer encounter the seed photons head-on (rear-on), i.e., at
superior (inferior) conjunction. Inversely, VHE absorption due to pair
production will be maximum (minimum) at superior (inferior)
conjunction. The phases of minimum and maximum flux in {\em Fermi} as
well as the anti-correlation with \hess\ are consistent with these
expectations, suggesting IC scattering is the dominant radiative
process above 100 MeV with the additional effect of pair production
above 30 GeV
\citep{Bednarek:2007qd,Dubus:2007oq,2008ApJ...674L..89S}. It is,
however, yet unclear whether the IC VHE emission is mainly produced
already in the pulsar wind zone of the system, given the high opacity
already found therein for particles accelerated at the pulsar site
\citep{2008APh....30..239S} or as a result of particle acceleration at
the shock formed in the wind collision region \citep{Dubus:2006lr} or
even well beyond the system \citep{2008A&A...489L..21B}.

However, the extension of these pictures from the TeV into the GeV
domain is undermined by the presence of an exponential cutoff at a few
GeV in the {\em Fermi} spectrum. A cutoff at $\sim$6.3 GeV was also
observed in \lsi\, indicating that this may be a common spectral
feature in this class of source. The companion star in \lsi\ is a Be
star with a dense equatorial disk. Passage of the compact object
through this disk might have explained the exponential cutoff: for
instance, this would crush a putative pulsar wind nebula closer to the
neutron star, increasing synchrotron losses and introducing a strong
dependence with orbital phase of the electron energy distribution
\citep{Dubus:2006lc}. But there is no such large, systematic contrast
in the density of the stellar wind from the O6.5V star along the orbit
in LS 5039. The presence of a similar cutoff in both systems argues
against explanations related to the properties of the orbit or
companion star.  The cutoff seems to require that the radiative process
or radiating electrons be different between the HE and VHE domains, in
disagreement with the picture presented above.

An intringuing possibility is that the emission in the {\em Fermi}
range from both LS 5039 and \lsi\ is magnetospheric emission as seen
in the dozens of pulsars that have now been detected by {\em
  Fermi}. The typical {\em Fermi} pulsar emission has a hard power-law
spectrum with a photon index $\approx$ 1.5 and an exponential cutoff
at $\approx 2.5$ GeV. In this case, the cutoff energy is thought to be
set by the balance between acceleration and losses to curvature
radiation. The emission should in such a case be pulsed, although the
orbital motion makes it exceedingly difficult to find in the {\em
  Fermi} data with no prior knowledge of the spin period. Pulsar wind
emission would dominate in the neighbouring hard X-ray and VHE bands,
supported by their similar orbital modulations
\citep{Hoffmann:2008ys,Takahashi:2008vu}.
It is to be noted that there is, however, an issue associated with
having dominant magnetospheric emission in the HE band: 
magnetospheric emission is usually thought to be due to
curvature radiation and has no obvious reason to be modulated with the
orbital motion (although magnetospheric emission
from pulsars in close binaries has hardly been modeled). 
The dense photon environment of the binaries may perhaps
introduce differences compared to gamma-ray magnetospheric emission
from isolated pulsars (for instance in the pair multiplicity).
Hence, the spectrum suggests a magnetospheric emission interpretation,
which is hard to reconcile with the IC scattering
interpretation suggested by the modulation.  Future HE and VHE should
lead to better constraints on the variability of the spectral
parameters along the orbit.  This will help resolve whether there are
several components and what their relative amplitudes are.

\acknowledgements
The \textit{Fermi} LAT Collaboration acknowledges generous ongoing
support from a number of agencies and institutes that have supported
both the development and the operation of the LAT as well as
scientific data analysis.  These include the National Aeronautics and
Space Administration and the Department of Energy in the United
States, the Commissariat \`a l'Energie Atomique and the Centre
National de la Recherche Scientifique / Institut National de Physique
Nucl\'eaire et de Physique des Particules in France, the Agenzia
Spaziale Italiana and the Istituto Nazionale di Fisica Nucleare in
Italy, the Ministry of Education, Culture, Sports, Science and
Technology (MEXT), High Energy Accelerator Research Organization (KEK)
and Japan Aerospace Exploration Agency (JAXA) in Japan, and the
K.~A.~Wallenberg Foundation, the Swedish Research Council and the
Swedish National Space Board in Sweden.

Additional support for science analysis during the operations phase is
gratefully acknowledged from the Spanish CSIC and MICINN, the Istituto
Nazionale di Astrofisica in Italy and the Centre National d'\'Etudes
Spatiales in France.

{\it Facility:} \facility{{\em Fermi}}

\end{document}